\documentclass{article}

\usepackage{arxiv}

\usepackage[utf8]{inputenc} 
\usepackage[T1]{fontenc}    
\usepackage{hyperref}       
\usepackage{url}            
\usepackage{booktabs}       
\usepackage{amsfonts}       
\usepackage{nicefrac}       
\usepackage{microtype}      
\usepackage{lipsum}
\usepackage{graphicx}
\graphicspath{ {./images/} }

\title{Evaluating the Impact of an Explainable Machine Learning System on Interobserver Agreement in Chest Radiograph Interpretation}

\author{
\textbf{Hieu H. Pham} \\ Coordinated Science Laboratory, UIUC \\ VinUni-Illinois Smart Health Center \& CECS, VinUniversity  \\ \texttt{hieu.ph@vinuni.edu.vn} \\
   \And
\textbf{Ha Q. Nguyen} \\ VinBigData JSC \\ \texttt{v.HaNQ3@vinbigdata.org} \\
  \And
 \textbf{Hieu T. Nguyen} \\ Northeastern University \\ \texttt{nguyen.trungh@northeastern.edu} \\
 \and
 \textbf{Linh T. Le} \\ Hanoi Medical University  \\ \texttt{linhdhyhn2017@gmail.com} \\
 \and
 \textbf{Khanh Lam} \\ 108 Hospital \\ \texttt{lamkhanh.himed@gmail.com} 
}

\begin{document}
\maketitle

\section{Introduction}
The actual impact of AI systems on the diagnostic performance of radiologists in clinical practice remains unclear \cite{seyyed2021underdiagnosis,larrazabal2020gender,seyyed2020chexclusion}. We developed an explainable deep learning system called VinDr-CXR that can classify a chest X-ray (CXR) into multiple thoracic diseases and localize critical findings on the image ~\cite{pham2021interpreting,nguyen2022deployment,tran2021learning,nguyen2022vindr,pham2022accurate,nguyen2022vindrp,le2023learning,nguyen2022learning,nguyen2021clinical}. A prospective study was conducted to measure the clinical impact of the VinDr-CXR in assisting six experienced radiologists. The results indicated that when VinDr-CXR was used as a diagnosis-supporting tool, significantly improved the agreement between radiologists themselves with an increase of 1.5\% in mean Fleiss' Kappa \cite{rucker2012measuring}. We also observed that, after the radiologists consulted VinDr-CXR's suggestions, the agreement between each of them and the system was remarkably increased by 3.3\% in mean Cohen's Kappa. This work has been accepted for publication in IEEE Access  and its full-length version can be found in \cite{pham2022accurate}. This is our short version submitted to the Midwest Machine Learning Symposium (MMLS 2023), Chicago, IL, USA (\url{https://www.midwest-ml.org/2023/}).

\section{Our Approach} 
The proposed framework includes two major components. First, an image-level classification network \cite{tan2019efficientnet} accepts a CXR scan as input and predicts whether it could be normal or abnormal. Second, a lesion-level detection network~\cite{tan2019efficientnet,tan2020efficientdet} receives an abnormal CXR scan as input from the classifier and provides the location of abnormal findings via bounding box predictions. The core of the VinDr-CXR system is based on state-of-the-art DL networks for image classification and object detection tasks. VinDr-CXR was trained on 51,485 CXR scans with radiologist-provided bounding box annotations \cite{nguyen2022vindr}. The actual impact of the VinDr-CXR was evaluated through a reader study (\textit{N} = 400). The inter-rater agreement among radiologists as well as the rate of agreement between VinDr-CXR and radiologists are then assessed the Cohen's Kappa metric.
\begin{figure}
\centering
\includegraphics[width=7cm,height=12cm]{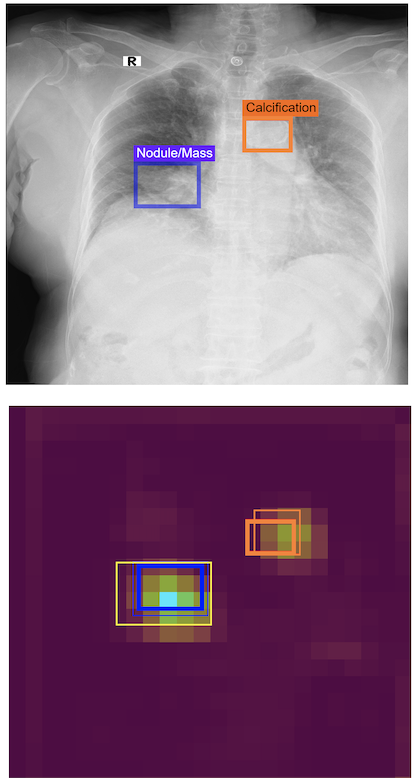}
\includegraphics[width=7cm,height=12cm]{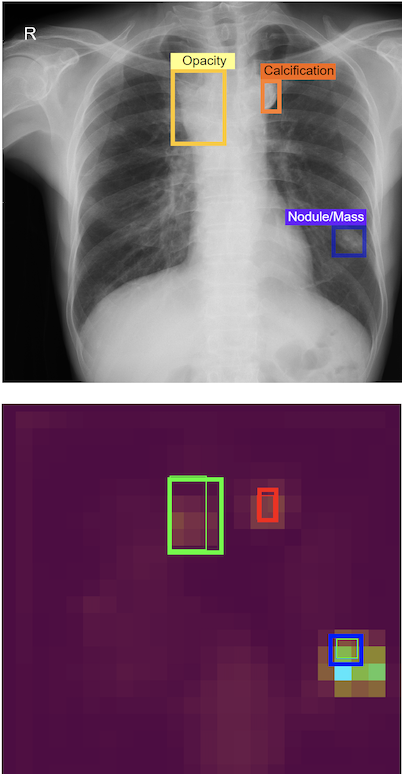}
\caption{VinDr-CXR localizes critical findings on the image.}
\label{fig:vis-xai}
\end{figure}
On a validation set of 3,000 CXR studies. The system reported a mean AUROC of 0.967 (95\% CI: 0.958, 0.975) for the classification task. For the detection, it achieved a sensitivity of 80.2\% (81.4, 84.9) at 1.0 false-positive marks per image. The FROC of the VinDr-CXR system was 78.36\% (76.46, 80.16).

\newpage 

\subsection{Impact of VinDr-CXR in clinical practice} 
 We recruited a group of six board-certified radiologists from 108 hospital (H108) and Hanoi Medical University (HMUH) to participate in our observer performance test. The reader study was conducted in two sessions. In the first session, participating readers read the CXR scans independently without the VinDr-CXR assistance. During the second session, the readers re-evaluated all CXR scans with the assistance of the VinDr-CXR. Specifically, the radiologists were provided the VinDr-CXR predictions in the form of bounding boxes, which locate abnormalities. They considered the model’s prediction and modified the diagnostics. Our experiments showed that in the second read, with the support of the VinDr-CXR system, agreement among three H108's radiologists was moderate with a Fleiss' Kappa of 0.545 (0.465, 0.625), corresponding to a 3.0\% improvement in  Fleiss' Kappa compared to the first read. Additionally, we found that the rate of VinDr-CXR agreement with the participating radiologists was slightly higher than the rate of agreement among radiologists. The agreement between each radiologist and the system was remarkably increased by 3.3\% in mean Cohen's Kappa. 
\begin{figure}
\centering
\includegraphics[width=15cm,height=5cm]{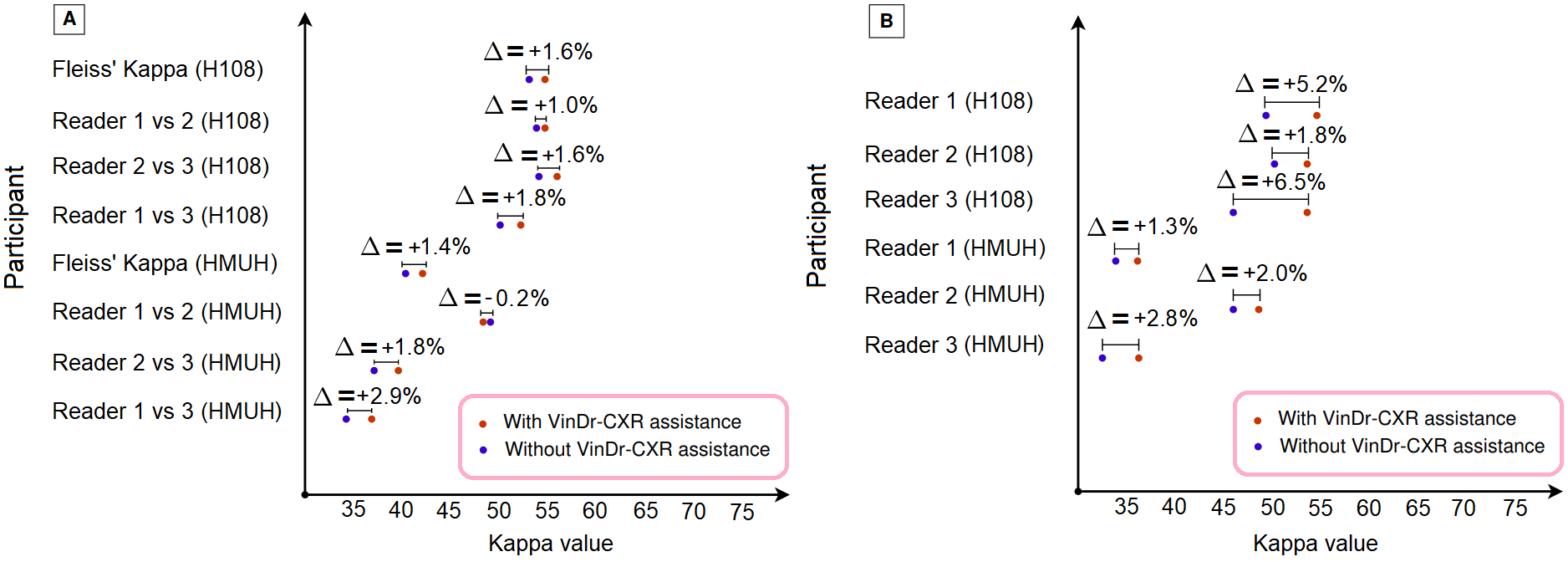}
\caption{Change in inter-radiologist agreement before and after consulting the VinDr-CXR predictions.}
\label{fig:reader_study}
\end{figure}

\newpage 

\section{Conclusion}
This study showed that an accurate and explainable deep learning system is able to improve interobserver agreement in the interpretation of chest radiograph. Further research is needed to validate the model prospectively and determine its utility in clinical settings. 

\bibliographystyle{abbrv}
\bibliography{references}

\end{document}